# Proton Irradiation Induced Defects in $\beta$-Ga$_2$O$_3$: a combined EPR and Theory Study

*Hans Jürgen von Bardeleben[1], Shengqiang Zhou[2], Uwe Gerstmann[3], Dmitry Skachkov[4], Walter R. L. Lambrecht[4], QuocDuy Ho[5] and Peter Deák[5]*

[1]Sorbonne Université, Institut des Nanosciences de Paris, UMR7588 au CNRS, 4 place Jussieu, 75005 Paris, France

[2]Helmholtz Zentrum Dresden Rossendorf, Institute of Ion Beam Physics and Materials Research, Bautzner Landstrasse 400, D-013128 Dresden, Germany

[3]Lehrstuhl für Theoretische Physik, Universität Paderborn, 33098 Paderborn, Germany

[4]Department of Physics, Case Western Reserve University, 10900 Euclid Avenue, Cleveland, OH-44106-7079, U.S.A.

[5]Bremen Center for Computational Materials Science, University of Bremen, P.O. Box 330440, D-28334 Bremen, Germany

**Abstract:**
Proton irradiation of both n-type and semi-insulating bulk samples of β-Ga$_2$O$_3$ leads to the formation of one paramagnetic defect with spin S=1/2, monoclinic point symmetry, a g-tensor with principal values of $g_b$=2.0313, $g_c$=2.0079, $g_{a*}$= 2.0025 and quasi isotropic superhyperfine interaction of 13G with two equivalent Ga neigbours. Its high introduction rate indicates it to be a primary irradiation induced defect. At low temperature, photoexcitation transforms this defect into a different metastable S=1/2 center with principal g-values of $g_b$=2.0064, $g_c$=2.0464, $g_{a*}$= 2.0024 and a reduced hyperfine interaction of 9G. This metastable defect is stable up to T=100K, when it switches back to the previous configuration. Density functional theory calculations of the Spin Hamiltonian parameters of various intrinsic defects are carried out using the Gauge Including Projector Augmented Wave method in order to determine the microscopic structure of these defects.Our results do not support the intuitive model of the isolated octahedral or tetrahedral gallium vacancy, $V_{Ga}^{2-}$, but favor the model of a gallium vacancy complex $V_{Ga}$-Ga$_i$-$V_{Ga}$.

**Introduction:**
β-Ga$_2$O$_3$ is a wide bandgap (4.8eV) semiconductor, which has been studied in the past by electron spin resonance (EPR) due to its interesting shallow donor properties. These early measurements, which were performed on non-intentionally doped n-type single crystals, concerned dynamic nuclear polarization and bistability effects [1,2]. At that time, the shallow donor was believed to be related to oxygen vacancy defects but recent theoretical predictions do exclude this model [3,4]. The EPR spectra of some 3d transition metals (Fe$^{3+}$,Mn$^{2+}$,Cr$^+$,Ti$^{3+}$) have equally been investigated [5-7]. Recently β-Ga$_2$O$_3$has attracted new interest due to its demonstrated applications in microelectronics and the availability of single crystals and epitaxial layers with controlled electronic properties. For a detailed review see reference [8]. Bothbulk single

crystals and doped epitaxial layers can now be purchased commerciallyfrom different suppliers

Intrinsic point defects in β-$Ga_2O_3$have not yet been clearly identified. They are expected to occur as native defects due to non stoichiometric growth conditions but can also be generated by irradiation with high energy particles. Very first EPR results of neutron irradiated [9], bulk samples have been published recently and the model of an octahedral gallium monovacancy defect has been proposed tentatively. Gallium vacancy defects and oxygen vacancy defects have also been evoked before in different optical and electrical studies [10-13] of irradiated samples but the assignment to a microscopic model is in generally not possible with those techniques. Very recently we have undertaken a detailed theoretical study [14] of various intrinsic point defects in β-$Ga_2O_3$ in order to identify the centers introduced by proton irradiation from their spin Hamiltonian parameters. In this work we present additional experimental results to further characterize these centers.

β-$Ga_2O_3$ has a monoclinic crystal structure with space group C2/m described by the three lattice vectors **a,b,c** and the angle β between **a** and **c** [15].This low symmetry structure gives rise to two nonequivalent gallium lattice sites and three nonequivalent oxygen lattice sites. Due to the different site symmetries, distorted octahedral and tetrahedral for the Ga sites and lower symmetry 3-fold and 4-fold bonded oxygen sites, a rather complicated situation can be expected. In addition as these intrinsic defects are deep centers, the Fermi level position is a key parameter,which determinestheircharge and spin states. In this wide bandgap material for most deep centers different charge transition levels occur in the gap [13] and thus the electronic configuration will change with electrical compensation. To be more specific, the two $V_{Ga}$ defects ($V_{Ga}$(tetra), $V_{Ga}$(octa))can take charge states from 0 to 3-; they are expected to be paramagnetic in the 2-, 1- and 0 charge states with a spin S=1/2, S=1, S=3/2and diamagnetic in the 3- charge state (fig1). Thus in n-type conductive samples the $V_{Ga}$ defects will not be observable by EPR. In Fe doped semi-insulating $Ga_2O_3$the tetrahedral $V_{Ga}$ becomes EPR active, while the octahedral one is still in the diamagnetic 3- charge state.

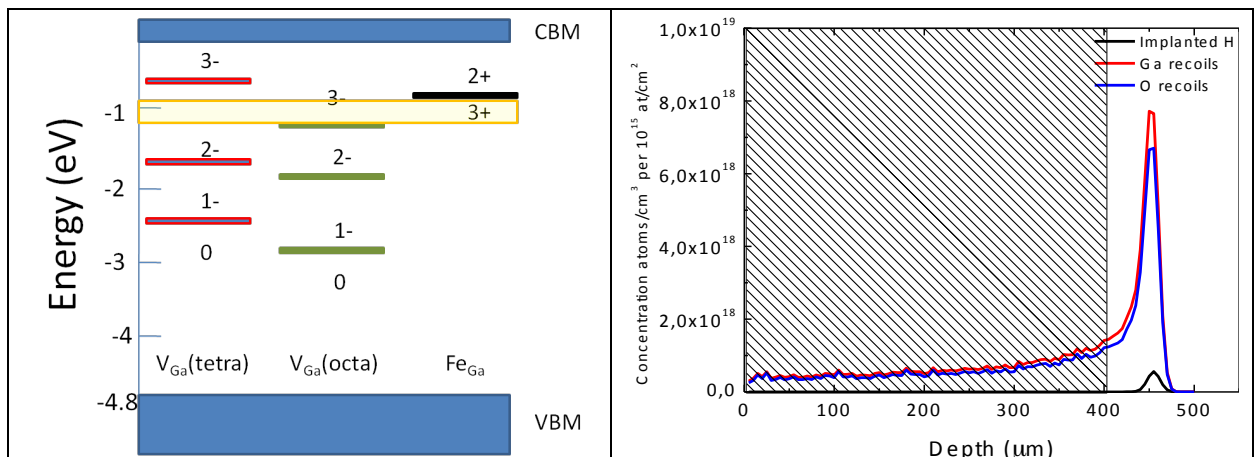

**Figure 1:** Charge transition levels for the gallium vacancies and the Fe acceptor [14]

**Figure 2:** Oxygen and Gallium vacancy distribution as predicted by SRIM simulations assuming displacement energies of 25eV for Gallium and 28eV for Oxygen atoms. Proton fluence: $1 \times 10^{15}$cm$^{-2}$.

In this work we further investigated the paramagnetic defects introduced in n-type and Fe doped semi-insulating samples by high-energy proton irradiation and modeled the spin Hamiltonian parameters by first principle calculations.

**Experimental:**

We investigated commercially purchased (Tamura) (010) oriented β-$Ga_2O_3$ bulk samples of 400μm thickness and 10x15$mm^2$ dimensions. The samples have been grown by the edge–defined film-fed (EFG) growth. Both non-intentionally doped n-type samples with a carrier concentration of $2x10^{17}cm^{-3}$ and Fe doped semi-insulating samples have been studied. Before irradiation the n-type samples displayed at T=300K only the spin S=1/2 EPR spectra of a shallow donor and the semi-insulating one's the spin S=5/2 spectra of the $Fe^{3+}$ impurities on the tetrahedral and octahedral lattice sites.

Both the initially n-type and the semi-insulating samples have been irradiated at room temperature with 12MV protons to a fluence of $10^{16}$ $cm^{-2}$. SRIM simulations of the irradiation induced vacancy formation (Fig.2) assuming displacement energies of 25eV (Ga) and 28eV (O) predict an introduction rate of the order of 500$cm^{-1}$ and place the end of range region outside the samples. Thus hydrogen related defects will not be considered in the following. Nevertheless, the displacement energies might well depend on the lattice sites and thus SRIM simulations are expected to give order of magnitude values only.

The EPR measurements have been performed with a CW X-band spectrometer in the temperature range from T=4K to 300K. The samples were measured under thermal equilibrium conditions and under in-situ optical excitation with light sources in the visible or ultraviolet region.

Absolute spin concentrations were determined with a calibrated spin standard sample ($Al_2O_3$:Cr) purchased from the National Bureau of Standards.

The g-tensors and hyperfine interactions of the main intrinsic defects have been obtained by first principle calculations. For details of the calculation procedures see ref.[14]. We calculated in particular the gallium vacancy and oxygen interstitial related defects, which were considered to be the most probable candidates for the paramagnetic centers observed our study.

**Experimental and calculation Results:**

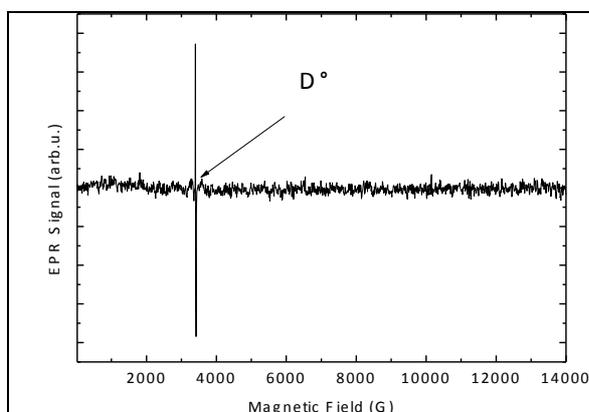 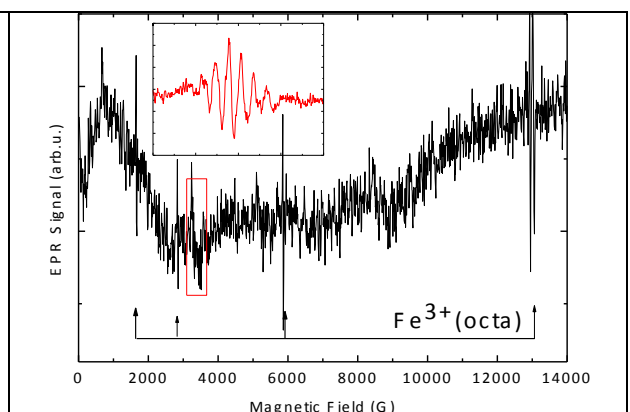

| **Figure 3**: Large scale EPR spectrum of the n-type sample before irradiation showing only the shallow donor resonance D° | **Figure 4**: Large scale EPR spectrum of the n-type sample after irradiation; the insert shows the EPR1 spectrum at higher resolution; B//**b** |

In Fig.3 we show typical room temperature X-band EPR spectra for an n-type sample before and after irradiation; the orientation of the applied magnetic field is B//**b** axis. Before irradiation only this EPR spectrum related to the neutral shallow donor is observed. It is characterized by a spin S=1/2 state, an anisotropic g-tensor, and a very narrow linewidth of $\Delta B_{pp}$=0.5G, which varies with the donor concentration. The shallow donor EPR spectra are identified from their monoclinic g-tensor with values g//**b**= 1.9630 [16]. The semi-insulating Fe doped samples show only two distinct EPR spectra with S=5/2previously assigned to $Fe^{3+}$[6]. The $Fe^{3+}$is a gallium substituted defect, occupying both tetrahedral and octahedral Ga lattice sites. It has a high spin S=5/2 ground state and gives rise in X-band to both allowed and forbidden transitions as the zero-field splitting parameter is of comparable magnitude to the Zeeman energy. Their EPR spectraare simplified at Q-band frequencies, where the $Fe^{3+}$ centers show the "classical" 5 line spectra in the 3000G to 19000G field range corresponding to the allowed $\Delta m_s$=+/-1 transitions. Their spin Hamiltonian parameters have been reported before [6].

After the proton irradiation (Fig.4) the initially n-type samples show no longer the shallow donor resonance but display the $Fe^{3+}$(octa) spectrum and an irradiation induced S=1/2 spectrum (EPR1), which is the object of this study. As might have been expected, the proton irradiation has electrically compensated the n-type samples due to the formation of deep acceptor centers [11]. This leads to a shift of the Fermi level towards midgap. Fe is a common residual contamination of the bulk samples introduced during the high temperature growth. The observation of the $Fe^{3+}$ spectrum gives us an information about the Fermilevel position (Fig.1) after the irradiation. It has been recently shown by DLTS measurements [17], that the deep center E2 with a ionization energy of 0.78eV, observed in bulk EFG samples, is associated with a Fe contamination. No distinction between $Fe^{3+}$(octa) and $Fe^{3+}$ (tetra) has been made in that study. The proton irradiation has thus moved the Fermilevel from the shallow donor position Ec-0.04eV to below Ec-0.78eV. Due to the lower Fermilevel position the EPR spectrum of the neutral shallow donor is no longer observed, as the donors will be ionized and thus diamagnetic (Fig.1).

In figure 5 we show a high resolution spectrum of the EPR1 center for B//**b**. The spectrum can be observed without a change of the spin Hamiltonian parameters between T=300K and 4K. It is characterized by an electron spin S=1/2, a monoclinic point symmetry, an anisotropic g-tensor with principal axes parallel to the crystal b,c,a* axes and a multiplet structure due to resolvedhyperfine interaction (table I). The simulation of the hyperfine structure shows that it is due to a superhyperfine interaction (SHF) with two equivalent Ga neighbors. Models assuming SHF interaction with 1,3 or 4 equivalent Ga neighbors or multiple nonequivalent Ga neighbors are not compatible with observed SHF structure. It is the presence of two Ga isotopes ($^{69}$Ga, $^{71}$Ga) with different nuclear moments and different isotopic abundance which gives rise to very characteristic HF patterns (Fig.6). The HF interaction with oxygen-if present-is in general not observable in samples non modified isotopically, due to the very low (0.038%) isotopic abundance of the isotope $^{17}$O, which is the only oxygen isotope with a nuclear spin (I=5/2); themain isotope $^{16}$O has no nuclear spin. The principal values and axes of the g-tensor have been determined by the measurement of the angular variation of the EPR spectrum for a rotation of the applied magnetic field in three lattice planes.The g-tensor is characterized by one g-value with a large deviation from the $g_e$=2.0023 and two values close to the free electron $g_e$-value (table I). Such a g-tensor is typical for a hole center on an oxygen p-orbital, the expected configuration of the Ga vacancy. Its values can be compared to the case of the zinc vacancy $V_{Zn}^{2-}$ in ZnO, which is equally a spin S=1/2 center [18]. In ZnO the two $V_{Zn}^{2-}$ centers are characterized by g-values of $g_{//}$=2.0024, $g_L$=2.0193 and $g_{xx}$=2.0173, $g_{yy}$=2.0183, $g_{zz}$=2.0028 for the axial and basal configurations. These values are quite similar to those observed for the EPR1 defect and point clearly to a gallium vacancy related defect model. However the situation in $Ga_2O_3$ is more complex due to the low symmetry of the Ga and O sites. In β-$Ga_2O_3$ even for an octahedral (tetrahedral) gallium vacancy the spin properties will in addition depend on which of the nonequivalent oxygen atoms the p-hole is localized. Oxygen hole centers, such as $O^-$ or $O_2^-$, have also been widely studied in the past in the ionic alkali halide compounds [19]. An oxygen hole center configuration might also occur for interstitial oxygen centers, another primary defect. We have thus considered this possibility also in our modeling. We have already shown in [14] that the oxygen interstitial is not stable and will form an $O_2^-$ center in the form of an oxygen dumbbell.

In order to evaluate the spin concentration of the EPR1 center we have measured it together with the $Al_2O_3$: Cr spin standard sample and evaluated the spin concentration by a double integration of the EPR spectrum. We obtain a defect concentration of $10^{18}$cm$^{-3}$, which is of the same order of magnitude as expected from the SRIM modeling for a primary defect. We conclude thus that the EPR1 center is a primary defect and its annealing stage is above room temperature.

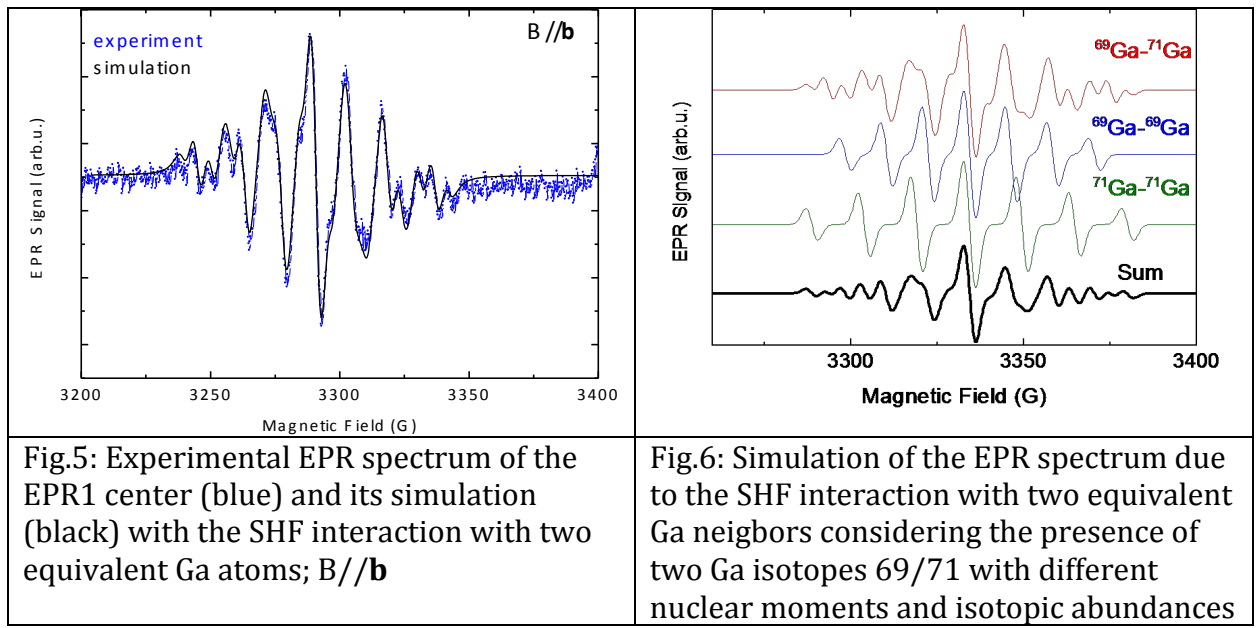

Fig.5: Experimental EPR spectrum of the EPR1 center (blue) and its simulation (black) with the SHF interaction with two equivalent Ga atoms; B//**b**

Fig.6: Simulation of the EPR spectrum due to the SHF interaction with two equivalent Ga neigbors considering the presence of two Ga isotopes 69/71 with different nuclear moments and isotopic abundances

After low temperature photoexcitation in the UV, the EPR1 spectrum is completely quenched and a new spin S=1/2 EPR spectrum (EPR2)(fig.7) with slightly modified Spin Hamiltonian parameters and comparable intensity is observed. The optical excitation has a threshold of 2.8eV (fig.8). The spin Hamiltonian parameters of the EPR2 center are given in table II. It has similar anisotropic g-values, but the principal axis of the highest g-value is shifted to the c-axis.

In the Fe doped semi-insulating samples we observe after proton irradiation at thermal equilibrium the same EPR1 center in addition to the two spectra of $Fe^{3+}$(octa) and $Fe^{3+}$(tetra). Under UV photoexcitation at temperatures below T<100K the same transformation to the EPR2 center is observed. The EPR2 center is metastable at low temperature up to T=100K.

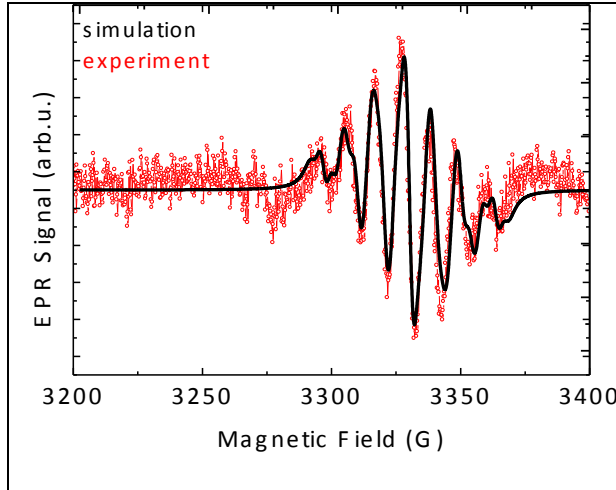

Fig.7: Experimental EPR spectrum of the EPR2 center (red) and its simulation (black) with the SHF interaction with two equivalent Ga atoms

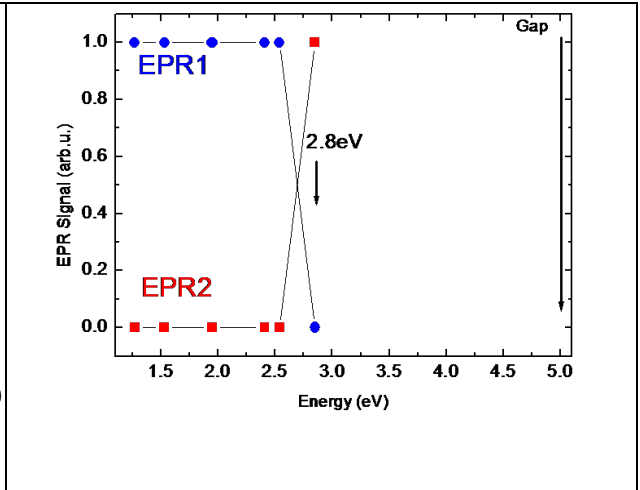

Fig.8: Spectral dependence of the optically induced generation of the EPR2 spectrum and the extinction of the EPR1 spectrum

The particular g-values of both centers EPR1 and EPR2 are characteristic for oxygen hole centers [14, 18, 19]. The principal axis associated with the g-value close to 2.0023 indicates in the simple model of a p-hole the p-orbital orientation [18,19]; within this model the p-orbital is directed along the crystal **a**-axis; it is not modified for the EPR2 centerand stays parallel to the **a**-axis for both centers. Only the principal axis associated with the highest g-values is now switched from **b** to the **c**-axis. The SHF interaction of the EPR2 center is slightly reduced, but is still characterized by the interaction with two equivalent Ga neighbors.

|  | Spin S | O atom | $g_b$ | $g_c$ | $g_{a*}$ | SHF N (Ga) | A ($^{69}$Ga) (G) |
|---|---|---|---|---|---|---|---|
| **Experiment EPR1** | **1/2** | **-** | **2.0313** | **2.0079** | **2.0025** | **2** | **13.8 14.6 12.8** |
| Model $V_{Ga}^{2-}$ (octa) | 1/2 | O(2) | 2.0258 | 2.0085 | 2.0184 | 2 Ga(tetra) | -22 |
| Model $V_{Ga}^{2-}$ (tetra) | 1/2 | O(1) | 2.0242 | 2.0068 | 2.0198 | 2 Ga(octa) | -22 |
| $V_{Ga}$(tetra)-$Ga_i$-$V_{Ga}$(tetra) | 1/2 | O(1) | 2.0251 | 2.0147 | 2.0048 | 2 | -21 |

**Table I**: experimental and calculated EPR parameters for the EPR1 center and the undistorted 2- charged Ga monovacancies and the $V_{Ga}$-$Ga_i$-$V_{Ga}$ complex; **S** theelectron spin, **O** the oxygen atom, on which the p-hole is localized, **g** the principal g-values, **N** the number of interacting Ga neigbors and their site symmetry , and **A** the SHF interaction parameter

To interpret the photo-induced transformation from EPR1 to EPR 2 we have considered two models: (i) EPR2 is a metastable configuration of the EPR1 center, and the metastable state is separated from the ground state by a barrier of the order of 0.1eV; (ii) the optical excitation corresponds to a charge transfer from the defect EPR1 to a second different defect EPR2. As the spin Hamiltonian parameters are rather similar, the microscopic structure of the EPR1 and EPR2 centers must be equally very similar.

To get further insight in the microscopic structure of the EPR1, EPR2 defects, we have calculated the Spin Hamiltonian parameters (g-tensor, SHF) for various defect models, including of course the simple models of a 2- charged gallium monovacancy at a tetrahedral or octahedral site (fig.9,10); they are the most obvious candidates for these two centers.

Concerning the center EPR1: whereas the g-tensor anisotropy and the SHF interaction with two Ga neighbors is reproduced by both models $V_{Ga}^{2-}$(tetra), $V_{Ga}^{2-}$(octa), (table I) the numerical values do not match the experimental ones and thus these two models must be discarded. We have equally calculated the parameters of a Ga vacancy complex model, proposed by Varley et al [20]; they predicted the tetrahedral Ga monovacancy $V_{Ga}^{2-}$(tetra) not to be stable and to transform into a $V_{Ga(tetra)}$-$Ga_i$-$V_{Ga(tetra)}$ complex. The spin Hamiltonian parameters for this model (Fig.11), shown in table I, are in much better agreement with the experimental results and thus this model is considered as the best candidate for the EPR1 center.

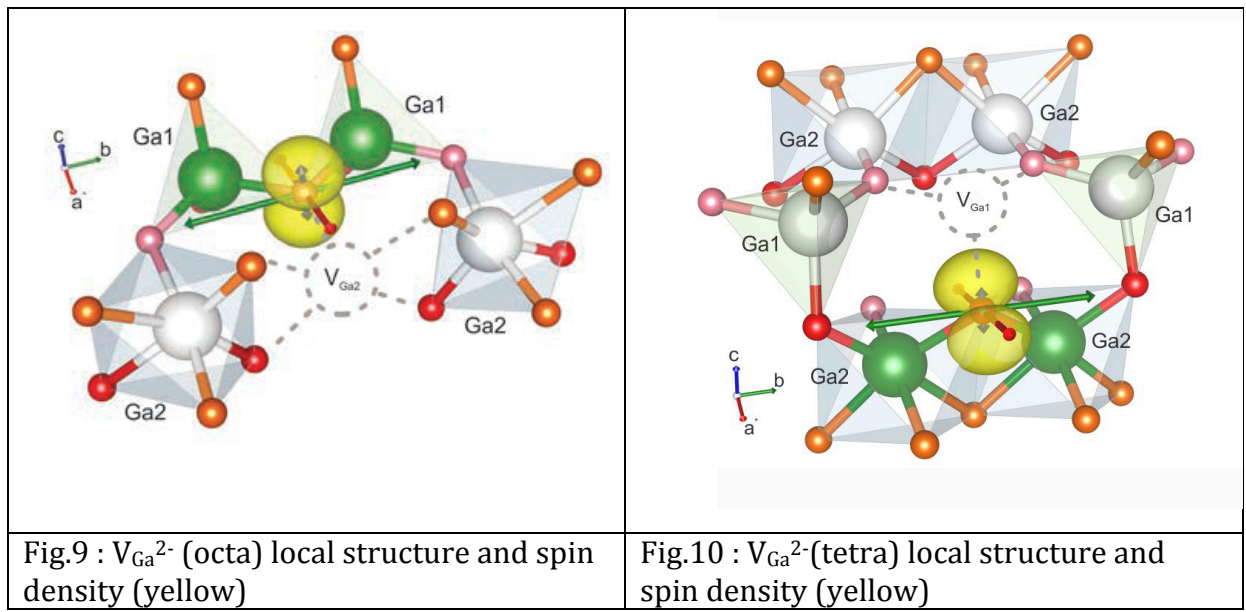

| Fig.9 : $V_{Ga}^{2-}$ (octa) local structure and spin density (yellow) | Fig.10 : $V_{Ga}^{2-}$(tetra) local structure and spin density (yellow) |
|---|---|

Concerning EPR2: as by modeling we have not found an optically excited metastable state for the $V_{Ga}$ complex center, we consider the EPR2 center as a different defect. Its observation after photoexcitation must be due to a charge transfer process to a different primary center. As this center is not observed in thermal equilibrium it must have been in a different and diamagnetic charge state. We have calculated the EPR parameters of various models for comparison with the EPR2 center; in particular we calculated the properties of a metastable octahedral Ga vacancy $V_{Ga}(octa)'$ (Fig.12), a self trapped hole (STH) center and an oxygen interstitial $O_i$ forming a $O_2^-$ dumbbell. It should be noted that the EPR2 center has previously been observed by Kananen et al [21] in n° irradiated samples after low temperature X-ray excitation; these authors attributed ad-hoc the spectrum to a self trapped hole center (STH) without detailed modeling. Our calculations do not support this assignment, as the calculated g-values do not agree with the experimental findings (table II). A best agreement is obtained for the oxygen dumbbell model, which reproduces nicely the g-tensor properties, but indicates a SHF interaction with (2+1) Ga neighbors. However, the simulation of the spectrum shape in the case of SHF interaction with three (2+1) Ga neigbors changes the structure drastically and is not compatible with the experimental results. Thus this model must also be discarded.

|  | Spin S | O atom | $g_b$ | $g_c$ | $g_{a*}$ | N (Ga) | A ($^{69}$Ga) (G) |
|---|---|---|---|---|---|---|---|
| **experiment** | **1/2** | **?** | **2.0064** | **2.0464** | **2.0024** | **2** | **9.8** **9.4** **9.0** |
| Model STH | 1/2 | O(1) | 2.0228 | 2.0237 | 2.0113 | 2 1 | -8 -16 |
| $V_{Ga}$(oct) metastable | 1/2 | O(1)O(1) | 2.0183 | 2.0372 | 2.0203 | 2 2 | -16 -21 |
| $O_2^-$ dumbbell | 1/2 | O(1)+$O_i$ | 2.0060 | 2.0306 | 2.0034 | 2 1 | -8 -19 |

**Table II**: experimental and calculated EPR parameters for the EPR2 center and the STH center, the O dumbbell and a distorted octahedral Ga monovacancy; **S,** the electron spin, **O,** the oxygen atom, on which the p-hole is localized, **g** the principal g-values, **N** the number of interacting Ga neigbors and their site symmetry, and **A** the SHF interaction parameter

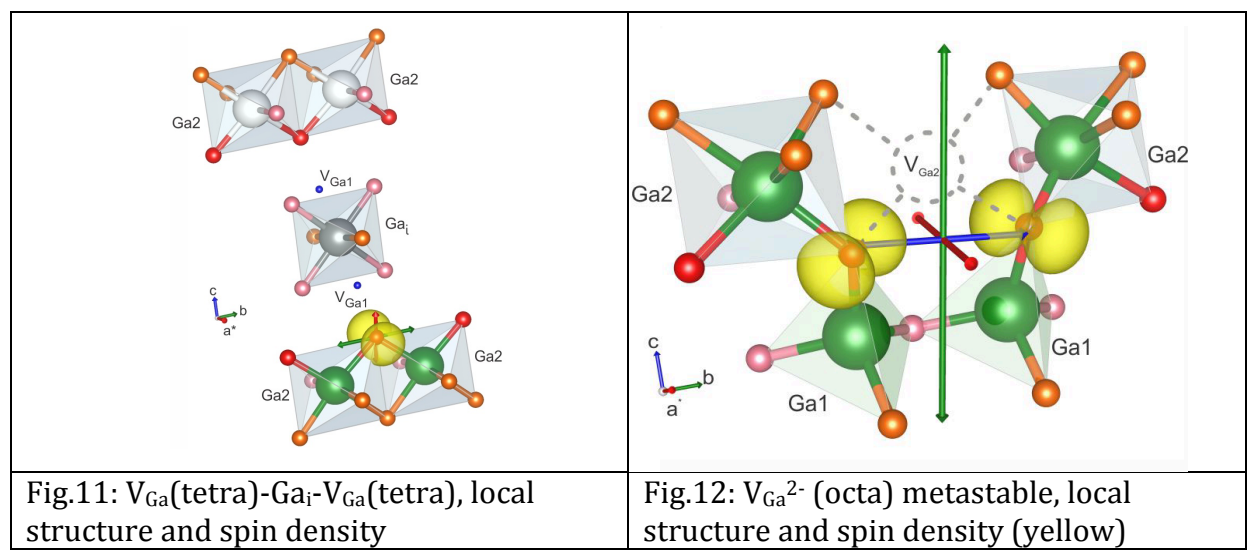

| Fig.11: $V_{Ga}$(tetra)-$Ga_i$-$V_{Ga}$(tetra), local structure and spin density | Fig.12: $V_{Ga}^{2-}$ (octa) metastable, local structure and spin density (yellow) |

## Conclusion:

Proton irradiation introduces two paramagnetic defects in β-Ga$_2$O$_3$, which are stable at room temperature. The high introduction rate shows them to be primary defects. Their g-tensor properties are characteristic for gallium vacancy centers but the numerical values are not compatible with the model of an undistorted gallium monovacancy on a tetrahedral or octahedral site (EPR1 center) or a self trapped hole center (EPR2 center). The most probable model for the EPR1 center is the V$_{Ga}$(tetra)-Ga$_i$-V$_{Ga}$(tetra) complex , the stableconfiguration of the V$_{Ga}^{2-}$(tetra) center. None of the considered models matches the parameters of the EPR2 center perfectly. Considering the Fermilevel dependence of the charge transition levels 3-/2-, the two vacancies V$_{Ga}$(tetra), V$_{Ga}$(octa) in the modified configurations are still the best candidates for the modeling of the EPR1,EPR2 defects.

## Acknowledgments:
The workat CWRU was supported by the National Science Foundation under grant No. DMR-1708593. The calculations were carried out at the Ohio SupercomputerCenter. The workat BCCMS was supported by the DFG grant No. FR2833/63-1 and by the Supercomputer Center of Northern Germany (HLRN Grant No. hbc00027). U.G. acknowledges support by the Deutsche Forschungsgemeinschaft (DFG) priority program SPP-1601.The ion irradiation was done at the Ion Beam Center of Helmholtz-Zentrum Dresden-Rossendorf. The authors like to thank Sh. Akhmadaliev for his assistance.